\documentclass[oneside]{article}
\usepackage{amsfonts}

{}
{}
{}
{}
{}
{}

\newcommand{\ra}{\rightarrow}

\title{{\bf Beyond Structured Programming}}
\author{M.H. van Emden\\
        {\small Technical Report DCS-359-IR}\\
        {\small Department of Computer Science}\\
        {\small University of Victoria}
       }
\date{}

\begin{document}
\maketitle
\date{}

\begin{abstract}
The correctness of a structured program is, at best,
plausible.
Though this is a step forward compared to what came before,
it falls short of verified correctness.
To verify a structured program according to Hoare's method
one is faced with the problem of finding assertions
to fit existing code.
In 1971 this mode of verification was declared by Dijkstra
as too hard to be of practical use---he advised that proof and
code were to grow together.
A method for doing this was independently
published by Reynolds in 1978 and by van Emden in 1979.
The latter was further developed to attain the form of
matrix code.
This form of code not only obviates the need of fitting
assertions to existing code, but helps in discovering an
algorithm that reaches a given postcondition from a fixed
precondition.
In this paper a keyboard-editable version of matrix code
is presented that uses E.W. Dijkstra's guarded commands as
starting point.
The result is reached by using Floyd's method rather than
Hoare's as starting point.
\end{abstract}

\section{{\large Introduction}}\label{sec:intro}


Structured Programming is the name of a method introduced
by E.W. Dijkstra in 1969 \cite{dijkstra249}.
At the time the method was revolutionary:
it advocated the elimination of the goto statement
and the exclusive use of conditional, alternative,
and repetitive clauses.
In a few years structured programming evolved from controversy
to orthodoxy.
C.A.R. Hoare invented a logic-based verification method for
structured programs \cite{hoareAxiomatic}.

In the late 1960's Dijkstra developed the conviction
that the looming ``software crisis'' could only be averted
by formally verifying all code.
At the time Hoare's method was the only one available.
It was found difficult to use in practice.
Dijkstra maintained his conviction that code needed to
be verified.
He defended it in his 1971 paper
``Concern for correctness as guiding principle in program
composition'' \cite{dijkstra71}.
In this paper Dijkstra advocated developing the code and its
correctness proof in parallel.
He did not include any suggestions as to how this might
be done.
I accept concern for correctness as guiding principle
in program composition.
In this paper I review the first two stages of a
published method that acts on this concern and
present a recent, so far unpublished, third stage.

The method is based on Floyd's
verification method for flowcharts \cite{floyd67}.
To verify such programs,
assertions are attached to suitably selected
labels in the code. These assertions have to be such that
they are true of the computational state whenever execution
reaches the associated label.
This property is assured by the way
the flowchart connects the assertions
by tests or assignment statements.
The connection is expressed by what Floyd calls
``verification conditions'',
which are triples written by Floyd as $V_C(p;q)$
where $C$ is a test or assignment statement, $p$,
the ``precondition'' is an assertion attached to the input node
of $C$, and
the ``postcondition'' is an assertion attached to the output node
of $C$.
Hoare wrote $p \{C\} q$ instead of $V_C(p;q)$.
Most subsequent researchers wrote
$\{p\} C \{q\}$ and called it ``Hoare triple''.
In this paper I will use the latter notation and
refer to the triple as ``verification condition''.

Floyd left some things unsaid concerning
the connection between logic and code.
With the missing connection in place,
his idea deserves the name ``Floyd Logic''.
He gave only the application to flowcharts, although its scope
is wider.
This paper explains the wider scope and defines
the nature of Floyd Logic.

Floyd's method applies to flowcharts;
Hoare's method is formulated in terms of the control
primitives favoured by
the ``structured programming'' proposed by Dijkstra and widely
embraced.
This style of programming results in more readable code.
Of course readable code is preferable to code
that is unreadable without having any redeeming properties.
But, without a formal Hoare-style proof,
the correctness of a structured program is, at best, {\sl plausible}.
If we are more ambitious and want verified code, then
a structured program presents us with complete code
for which suitable assertions need to be found.

Such a situation was considered by Dijkstra in a 1971
paper \cite{dijkstra71} from which I quote
\begin{quotation}
{\sl
When correctness
concerns come as an afterthought and correctness proofs have to
be given once the program is already completed, the programmer
can indeed expect severe troubles. If, however, he adheres to the
discipline to produce the correctness proofs as he programs along,
he will produce program and proof with less effort than programming
alone would have taken.
}
\end{quotation}

Let us call this ``Dijkstra's Principle''.
Here he implicitly enjoined his readers to find an alternative
to Structured Programming.
But about the form this alternative is to take, nothing more
specific is to be found than that the programmer is to
``produce the correctness proofs as he programs along''.
In a later paper \cite{dijkstra75} he advocated
\begin{quote}
{\sl
$\ldots$ challenging the choice
of {\rm a posteriori} verification of given programs
as the most significant
problem, the argument against that choice being, that programs
are not ``given'' but must be designed. Instead of trying to find
out which known proof-patterns are applicable when faced with a
given program, the program designer can try to do it the other way
round: by first choosing the proof-pattern that he wants to be
applicable, he can then write his program in such s way as to satisfy
the requirements of that proof.
}
\end{quote}

The first alternative that might fit Dijkstra's suggestions
in \cite{dijkstra71,dijkstra75} was proposed,
according to R-J. Back \cite{back09},
by J. Reynolds \cite{reynolds78}
and by M.H. van Emden \cite{vanemden79}.
Van Emden's method, ``programming with verifications'',
was developed into the Matrix Code of 2014 \cite{vanemden14}.

To balance the generalities considered so far,
let us look at a concrete example of Programming with Verification
Conditions in the form of
the program in Figure~\ref{prog:spaghetti}.
\begin{figure}
\begin{center}
\begin{minipage}[t]{3in}
\hrule \vspace{2mm}
\begin{verbatim}
int gcd(int x, int y) {
S: // x=x0 & y=y0 & x0>0 & y0>0
  int z = 1; goto A;
A: // gcd(x0,y0) = z*gcd(x,y)
  if (x == y) { z *= x; goto H; }
  if (x > y) { goto B; }
  if (x < y) { swap(&x, &y); goto B; }
  assert(0);
B: // A & x>y
  if (y == 1) goto H;
  if (y > 1) goto E;
  assert(0);
E: // B & y>1
  if (x%2 == 0) goto C;
  if (x%2 != 0) goto D;
  assert(0);
C: // B & even(x)
  if (y%2 == 0) {x /= 2; y /= 2; z *= 2; goto B;}
  if (y%2 != 0) {x /= 2; goto A;}
  assert(0);
D: // B & odd(x)
  if (y%2 == 0) { y /= 2; goto B; }
  if (y%2 != 0) { x = x/2-y/2; goto A; }
  assert(0);
H: // z = gcd(x0, y0)
  return z;
}
\end{verbatim}
\hrule
\end{minipage}
\end{center}
\caption{
\label{prog:spaghetti}
Not your ordinary spaghetti code:
every label is associated with an assertion.
Each {\tt goto} statement is to be read as a claim
that the computational state satisfies the assertion associated
with the label.
The program can be read as a collection of verification
conditions.
Moreover, it can be seen that
execution does not traverse a closed circuit through
the code without decreasing $|x-y|$.
As the verification conditions are all true, it follows that,
when label H is reached, {\tt z} has the desired value.
}
\end{figure}
The purpose of the function is to compute the greatest
common divisor in time that is logarithmic in the greatest
of its arguments.
The coding style, or, rather, the absence of any,
may well leave readers at a loss for words.
I present the listing here to make the point that readability
in the usual sense is not relevant, as we should be
aiming at {\sl verifiability} rather than mere readability.

And verified it is:
the function in Figure~\ref{prog:spaghetti} is verified in the
sense that it is proved that executing a call {\tt gcd(x, y)}
results in $gcd(x,y)$.
The proof consists in the verification that
whenever execution reaches a label, the assertion associated
with that label (the one shown as a comment) holds.
This follows from the truth of the verification
conditions (in the sense of Floyd \cite{floyd67})
listed in Figure~\ref{vc:stein}.
It is verified so easily because that it was written
with this goal in mind.
\begin{figure}
\begin{center}
\begin{minipage}[t]{3in}
\hrule \vspace{2mm}
\begin{verbatim}
{S} int z = 1 {A}
{A} x == y; z *= x {H} 
{A} x  > y {B} 
{A} x  < y; swap(&x, &y) {B} 
{B} y == 1 {H} 
{B} y  > 1 {E} 
{E} x%2 == 0 {C} 
{E} x%2 != 0 {B} 
{C} y%2 == 0; x /= 2; y /= 2; z *= 2 {B} 
{C} y%2 != 0; x /= 2 {A} 
{D} y%2 == 0; y /= 2 {B}
{D} y%2 != 0; x = x/2-y/2 {A} 
\end{verbatim}
\hrule
\end{minipage}
\end{center}
\caption{
\label{vc:stein}
Verification conditions for Figure~\ref{prog:spaghetti}.
}
\end{figure}

These verification conditions need to be read with some
indulgence. An intuitive guideline is that both tests and
assignment statements can be interpreted as binary relations
over states and can be composed
as binary relations with the symbol ``{\tt ;}''. 
Moreover, assertions can be interpreted as sets of states.
In the sequel these interpretations will be stated
more precisely.

The validity of the verification conditions relies on
equalities concerning the {\it gcd} function shown
in Figure~\ref{tex:stein}.
\begin{figure}
\begin{center}
\begin{minipage}[t]{3in}
\hrule \vspace{2mm}
\begin{eqnarray*}
gcd(x,x) &=& x \\
gcd(x,1) &=& 1 \\
gcd(x,y) &=& gcd(y,x) \\
gcd(x,y) &=& 2*gcd(x/2, y/2)
   \mbox{ if } even(x) \wedge even(y) \\
gcd(x,y) &=& gcd(x/2, y)
   \mbox{ if } even(x) \wedge odd(y) \\
gcd(x,y) &=& gcd(x, y/2)
   \mbox{ if } odd(x) \wedge even(y) \\
gcd(x,y) &=& gcd(\lfloor x/2\rfloor-\lfloor y/2\rfloor, y)
   \mbox{ if } odd(x) \wedge odd(y) \wedge x>y \\
\end{eqnarray*}
\hrule
\end{minipage}
\end{center}
\caption{
\label{tex:stein}
Properties used in Stein's algorithm
of the $gcd$ function on positive integers.
}
\end{figure}
The method of obtaining a logarithmic-time algorithm 
based on these properties is due to Stein 
and brought to my attention by reading \cite{stepMcJones}.

As truth is a property of each verification condition by itself,
their ordering in Figure~\ref{vc:stein} does not matter. 
In Floyd's verification method there is a close correspondence
between the code and the verification conditions.
To make this correspondence easy to see, I have placed verification
conditions with the same precondition next to each other.
This makes it easy to transcribe the set
of verification conditions into code.
It matters little whether the code is in the form of a flowchart 
(as in Floyd's  paper) or in the form of C code
(as in Figure~\ref{prog:spaghetti}).
In both cases the correspondence
between sets of verification conditions
and code is close---Figure~\ref{prog:spaghetti} can be compiled,
yet in a sense is a theorem of mathematics.
In case of trouble, one must realize that
errors are not uncommon in theorems
and their proofs. When some of these show by the code giving wrong
results or no results, we are lucky,
but we should not make the mistake
of trying to debug the code:
there must be an error in the theorem
or in its transcription---tracking that down is a more
productive activity than debugging code.

\section{{\large Structured Programming}}\label{sec:structProgr}

``Structured programming'' has been, and continues to be,
widely influential and, arguably, widely misunderstood.
It is therefore useful to examine how the term arose,
its various meanings, and how it became influential.

As Dijkstra recounts in his retrospective memorandum
EWD1308 \cite{dijkstra1308},
1968 was a turning point in his career.
At this point he realized that in the larger scheme of
things his two biggest successes in software system implementation,
the first Algol 60 compiler of 1960 and the recently completed 
THE operating system, should be regarded as mere ``agility
exercises''. He decided 
``to tackle the real problem of How to Do Difficult Things''.
To get started he wrote a memorandum with title
``Notes on structured programming'', completed August 1969.
By the time it was published in 1971 (\cite{daDiHo}),
it had already become influential via the grapevine. 

In 1968 Dijkstra had submitted a short note to the Communications
of the ACM under the title ``A case against the goto statement''.
To expedite publication the editor turned it into a Letter to the
Editor. Such letters do not carry a title and are supplied by the
editor with a phrase summarizing the content.
Thus it appeared under the heading ``Goto Considered Harmful''
\cite{dijkstra68}.
Although Dijkstra did not discuss or even mention structured
programming, the combined effect of the Letter to the Editor and
``Notes on structured programming'' was that structured programming
was equated to abstaining from the use of the goto statement.
In 2001 Dijkstra wrote \cite{dijkstra1308}
``IBM $\ldots$ stole the term 'Structured Programming'
and under its auspices Harlan D. Mills trivialized the
original concept to the abolishment of the goto statement.''
Indeed a far cry from the intention of structured programming,
but who could have guessed it was meant as a step toward
``learning How to do Difficult Things''.

I sense that Dijkstra's sense of urgency was driven by two
considerations: (1) the rapid advance in computer hardware will
present the opportunity to build systems requiring millions
of lines of code and (2) without drastic methodological measures
such systems will never be made to perform reliably.
Dijkstra's conclusion was that structured programming would not
be enough; only a correctness proof would do.
He realized that, for example, Hoare's method of inserting
assertions into a structured program is too hard to be practically
usable and concluded \cite{dijkstra71} that correctness proof would
only be feasible if the code was designed {\sl ab initio} to
be easily proved correct. The question of abstention from the
goto statement did not come up. Correctly so: structured 
programming aims at readability; verifiability is a different,
independent, and more important goal.

While Dijkstra was distressed by the trivialization of
structured programming, C.A.R. Hoare reports \cite{hoare96} with
equanimity the fate of the concept in the decades following:
\begin{quote}
The decisive breakthrough in the adoption of structured programming
by IBM was the publication of a simple result in pure programming
theory, the Bohm-Jacopini theorem. This showed that an arbitrary
program with jumps could be executed by an interpreter written
without any jumps at all; so in principle any task whatsoever can
be carried out by purely structured code. This theorem was needed
to convince senior managers of the company that no harm would come
from adopting structured programming as a company policy; and project
managers needed it to protect themselves from having to show their
programmers how to do it by rewriting every piece of complex spaghetti
code that might be submitted. Instead the programmers were just
instructed to find a way, secure in the knowledge that they always
could. And after a while, they always did.
\end{quote}
One shudders to think of the contortions to which the code 
was subjected so that programmers could satisfy project managers:
``Look! No goto's!''.
But Hoare's intention was to convey the impression that
structured programming was responsible for more than
cosmetic changes.

\section{{\large Floyd Logic}}\label{sec:floydLogic}
Both Floyd \cite{floyd67} and Hoare \cite{hoareAxiomatic}
use the same identifiers for
variables in assertions as for variables in program code.
Apparently they do not consider this double use worth even a passing
remark, whereas it seems to some readers not obvious that there
is any connection between the variables of logic
(the things one quantifies over, for example)
and the variables in a programming
language (names for memory locations).
Whatever connection Floyd \cite{floyd67}
and Hoare \cite{hoareAxiomatic} had in mind was obviated by Apt \cite{tenYrs}
where the need to make
such connection is avoided by creating a new logic that included among
its expressions some that can be recognized as programming language
constructs.
In this paper I follow Floyd and Hoare in leaving the language
of assertions unchanged at first-order predicate
logic.
I therefore have to establish the connection between program
variables and logic variables.
I do this by defining the meaning of assertions and program
fragments in a suitable way.

\paragraph{The meaning of an assertion is a set of states\\}
Each assertion is 
a formula $F$ of first-order predicate logic.
For each free variable in $F$ there
is a program variable with the same identifier.

The collection $A$ of assertions of a program
have a common vocabulary,
which consists of function symbols,
predicate symbols, and a common set $V$ of variables.

An interpretation for $A$ consists of a universe of
discourse $D$, an assignment of relations over $D$ to the predicate
symbols, and an assignment of functions over $D$ to the function
symbols.
The truth $M^I_s(F)$ of a formula $F$
depends not only on an interpretation $I$,
but also on a {\sl state}\footnote{
In some logic texts a ``valuation''.
}$s$, which is a mapping of type $V\ra D$ of each variable in $V$
to an element of $D$.
Hence a state is a vector of elements of $D$ indexed by
elements of $V$.

The meaning $M^I(F)$ of a formula $F$ in an interpretation $I$
is the set of states in which it is true under that
interpretation:
$M^I(F) = \{s\in (V\ra D) \mid M_s^I(F)\}$
In other words, the meaning is a relation consisting of
tuples of type $V\ra D$.
In other words, the meaning is a subset of $V\ra D$.

Typically, the set $V_F$ of free variables in $F$ is a strict
subset of $V$.
Then $s\in M^I(F)$ implies that $s'\in M^I(F)$
for any $s'$ that only differs from $s$ in variables not in $V_F$.

Truth or falsity of a formula is determined
by an interpretation and a state individually for this formula.
In Floyd's logic the unit of interpretation is extended to be
the collection $A$ of assertions of the same program.
This is the only difference between Floyd's logic and
first-order predicate logic.
It is a difference because $A$, although consisting of formulas
of first-order predicate logic, is not itself a sentence or
formula of this logic. 

\paragraph{The meaning of a program is a
  binary relation over states\\}
By ``program'' I understand anything that can be characterized
as a binary relation over the set of states;
hence as a set of pairs of the form
$\langle$state before, state after$\rangle$.

Let us first establish what to include under ``program''.
We might as well make it general, observing that
the structured programs of Hoare
\cite{hoareAxiomatic} are a special case
of the flowcharts of \cite{floyd67}, which are themselves a special
case of programs defined as follows.
\begin{enumerate}
\item
A formula $F$ of first-order predicate logic is a program.
The meaning of program $F$ is the binary relation
$\{\langle s,s\rangle\mid M^I_s(F)\}$,
which is
the set of pairs of equal
states $s$ in which the formula is true in $I$.
\item
``$v := t$'' is a program, where $v$ is a variable
and $t$ is a term, both ``variable'' and ``term''
in the sense of first-order predicate logic.
Let $M_s^I(t) \in D$ be the meaning of term $t$
in state $s$ according to Tarskian semantics.
Then the meaning of ``$v := t$'' is the set 
$\langle p, q\rangle$ of pairs of states
such that $q(v) = M_p^I(t)$ and $q(x) = p(x)$
if $x\in V$ is not the same variable as $v$.
\end{enumerate}

\paragraph{The meaning of a verification condition\\}
$\{p\}C\{q\}$
is true iff
for all $s$ and $t$ in $V\ra D$,
$
\langle s,t\rangle\in M^I(C) \wedge s\in M^I(p)
$ 
implies $t\in M^I(q)$.

\section{{\large Programming language}}
The listing of verification conditions in Figure~\ref{vc:stein}
is satisfactory for the verification of the program in
Figure~\ref{prog:spaghetti}.
But we want to avoid having to find verification conditions
for existing code.
As we will show, it is easier to write verification conditions
first and then write code that is verified by it.
This is achieved by Matrix Code.
However, its two-dimensional format makes it awkward to enter and
edit.
It helps to create an intermediate form in a text file that is editable
by keyboard.
This has to be so that it is easy to transcribe verification conditions
to it and such that it is easy to write compilable code from it. 
This is a way of following Dijkstra's Principle,
and the only way I know of.

For this intermediate text
the {\tt if...fi} statement of Dijkstra's guarded-command
language \cite{dijkstra76} is a good starting point.
We make two modifications.
The first is to attach a label to each {\tt if...fi}
and to associate an assertion with this label.
The second modification is to append
{\tt goto L} at the end of each guarded command to claim
that the assertion labelled by {\tt L} is true when execution
reaches that point.
In this way each guarded command closely mirrors a verification
condition and is easy to transcribe into compilable code.

Guards are written as Algol boolean expressions,
hence ``{\tt =}'' for equality. The assignment operator is written
as ``{\tt :=}''.
Text between a label and an {\tt if} is comment.
It is used for the assertion associated with the label.
The Pascal convention of {\tt \{...\}} is used for comments.

As is the case in Dijkstra's guarded-command language,
{\tt if...fi} where all guards are false
is equivalent to {\tt abort}. This command can be added
after every {\tt fi} without altering the meaning,
so is normally not written.
It is useful when a replacement for {\tt if...fi}
is needed.

I call this language ``Liffig'' because it symbolizes an
{\tt if...if} preceded by a label, which is where the L comes from,
and is followed by a goto, which is where the g comes from.
The Liffig text to serve as intermediate between
Figure~\ref{vc:stein}
and
Figure~\ref{prog:spaghetti}
is in
Figure~\ref{liffig:stein}.
\begin{figure}
\begin{center}
\begin{minipage}[t]{3in}
\hrule \vspace{2mm}
\begin{verbatim}
S: x=x0 & y=y0
   if true -> z := 1; goto A
   fi
A: gcd(x0,y0) = z*gcd(x,y)
  if x=y -> z := z*x; goto H
   | x>y -> goto B
   | x<y -> swap(x, y); goto B
  fi
B: A & x>y
  if y=1 -> goto H
   | y>1 -> goto E
  fi
E: B & y>1
  if even(x) -> goto C
   | odd(x)  -> goto D
  fi
C: B & even(x)
  if even(y) -> x, y, z := x/2, y/2, 2*z; goto B
   | odd(y)  -> x := x/2; goto A
  fi
D: B & odd(x)
  if even(y) -> y := y/2; goto B
   | odd(y)  -> x := x/2 - y/2 ; goto A
  fi
H: gcd(x0,y0) = z
  return z
\end{verbatim}
\hrule
\end{minipage}
\end{center}
\caption{
\label{liffig:stein}
The Liffig program that is obtained by transcription from the
verification conditions in Figure~\ref{vc:stein}.
From this Liffig program
the compilable C code can be obtained by transcription.
I regard neither the verification conditions nor the
transcriptions to be in need of justification,
but may be in need of scrutiny for possible errors. 
The name ``Liffig'' is derived from the {\tt if...fi} construct of
E.W. Dijkstra's guarded-command language.
}
\end{figure}
I agree that it is presumptuous to dream up a name for such
a meagre parody of a language.
Yet, when one wants to talk about something
it helps to have a name for that something.
I can report that one gets used to ``Liffig''.

\section{{\large Program construction by gathering
  snippets of truth}}
Dijkstra's Principle is
``to produce the correctness proofs as he programs along''.
With a bit of good will, programming with verification conditions
can be regarded as a realization of the principle.
Here partial correctness is the key.
Partial correctness means that the program cannot do anything 
wrong. It is short of total correctness when it does not enough.
So one can always start by achieving partial correctness
with a program having the desired
precondition and postcondition that does nothing at all.
Successively larger parts of the problem
are solved by adding patently correct
increments in the form of true verification conditions.
The importance of programs and
assertions as defined here is that these allow us to follow the latter
approach.
In Liffig, programs can grow by adding verification conditions.
By choosing verification conditions with sufficiently simple
program components their truth is obvious.
Because of this simplicity, the verification condition
is a mere ``snippet of truth''.
Hence ``program construction by gathering snippets of truth'',
a useful slogan to characterize this realization of Dijkstra's
Principle.

\subsection{Example: GCD}
Let us follow Dijkstra's Principle to obtain the program in
Figure~\ref{liffig:stein}.
The snippets of truth are obtained from properties of
the $gcd$ function (Figure~\ref{tex:stein}),
starting with $gcd(x,x) = x$.
\begin{verbatim}
S: x=x0 & y=y0
  if true -> z := 1; goto A
  fi
A: gcd(x0,y0) = z*gcd(x,y)
  if x=y -> z := z*x; goto H
  fi
B: A & x>y
  abort
H: gcd(x0,y0) = z
  return z

\end{verbatim}
The partial correctness is very partial indeed:
we only get a successful computation when the arguments
are equal.

Using $gcd(x,y) = gcd(y,x)$ we get
\begin{verbatim}
S: x=x0 & y=y0
  if true -> z := 1; goto A
  fi
A: gcd(x0,y0) = z*gcd(x,y)
  if x=y -> z := z*x; goto H
   | x>y -> goto B
   | x<y -> swap(x, y); goto B
  fi
B: A & x>y
  abort
H: gcd(x0,y0) = z
  return z
\end{verbatim}
This does not solve a larger part of the problem,
but opens the way to more effective program increments,
such as the one that uses $gcd(x,1) = 1$.

\begin{verbatim}
S: x=x0 & y=y0
  if true -> z := 1; goto A
  fi
A: gcd(x0,y0) = z*gcd(x,y)
  if x=y -> z := z*x; goto H
   | x>y -> goto B
   | x<y -> swap(x, y); goto B
  fi
B: A & x>y
  if y=1 -> goto H
   | y>1 -> goto E
  fi
E: B & y>1
  abort
H: gcd(x0,y0) = z
  return z
\end{verbatim}

We continue adding increments in this manner
until we have used all of
the equalities in Figure~\ref{tex:stein}.
The result is the program in Figure~\ref{liffig:stein}.

\paragraph{Thoughts on the example}
In what sense does Figure~\ref{prog:spaghetti}
specify an {\sl algorithm}?
With the method demonstrated here it is merely the result
of the composition of two
transcriptions of the equalities in Figure~\ref{tex:stein}.
Of course the equalities are not a random selection from among the
infinity of equalities that are true of the $gcd$ function.
Rather than an algorithm, Stein's contribution consists of this
selection.

\subsection{Example: Fast exponentiation}
Exponentiation is often explained as iterated multiplication.
This suggests an algorithm for computing $a^n$ that uses
in the order of $n$ operations.
This is improved by using $a^n = (a^2)^{(n/2)}$ at appropriate
points in the algorithm.
In this way the number of operations can be reduced to
the order of $\log n$.

This example, a staple of introductions to programming,
has a wider significance as suggested by the fact that the same
idea turns up in surprising places.
Stepanov and McJones \cite{stepMcJones}
and Stepanov and Rose \cite{stepRose} explain this
wider significance by pointing out that $na$ can be
interpreted as 
$$
\underbrace{a+a+\cdots +a}_{n \: \mathit{times}}
$$
in any algebra that has a binary associative $+$ and
a neutral element.
This explains why the same programming trick works to
obtain multiplication from addition (``Egyptian multiplication''
in \cite{stepRose}), the computation of high powers in
modular arithmetic (used in cryptography),
and computation of the $n$-th Fibonacci number in logarithmic
time, to mention just a few examples.

For the purpose of this paper, fast exponentiation is a good
example because of different attitudes of different authors.
After the simplest version Stepanov and Rose dedicate a
section, 2.2, to ``Improving the algorithm'',
where they address various places where needless tests
or operations occur.
Dijkstra and Feijen \cite{dijkstraFeijen}
point out such possibilities,
but loftily dismiss them: ``All such attempts probably make the
program text less clear, and in any case much longer''.
In this paper clarity of program text is not a valid criterion,
verifiability is.
Section 2.2 of \cite{stepRose}
shows that improving the simplest, most elegant, version
is an instructive exercise, although opinions apparently differ
as to whether it is worthwhile.

In the remainder of this section we gather snippets
of truth in a way that results in an algorithm that seems
to be as efficient as the one developed in
section 2.2 of \cite{stepRose}.

We begin by documenting the precondition and the postcondition
of the entire program.
\begin{verbatim}
S: { n0 >= 0 & a = a0 & n = n0 & z = 0 }
H: {z = n0*a0}
\end{verbatim}
We consider whether there is sufficiently simple {\tt C}
such that \verb+{S}C{H}+.
As there is none,
we find an assertion that is a common generalization of
the precondition and the postcondition.
This is Dijkstra's ``proof pattern that he wants to be
applicable'' \cite{dijkstra75}.
\begin{verbatim}
A: n >= 0 & n0*a0 = z + n*a
\end{verbatim}
We pick off the easy cases $n=0$ and $n=1$.
\begin{verbatim}
S: n0 >= 0 & a = a0 & n = n0 & z = 0
if true -> goto A
fi
A: n >= 0 & n0*a0 = z + n*a
if n=0 -> goto H
 | n!=0 -> { n > 0 } goto B
fi
B: A & n>0
if n = 1 -> z := z+a; goto H
 | n != 1 -> { n > 1 } goto C
fi
C: A & n>1
"reduce n while maintaining A"
H: z = n0*a0
\end{verbatim}
To reduce $n$ we can decrement or divide by two,
with priority given to the latter.
\begin{verbatim} 
S: n0 >= 0 & a = a0 & n = n0 & z = 0
if true -> goto A
fi
A: n >= 0 & n0*a0 = z + n*a
if n=0 -> goto H
 | n!=0 -> { n > 0 } goto B
fi
B: A & n>0
if n = 1 -> z := z+a; goto H
 | n != 1 -> { n > 1 } goto C
fi
C: A & n>1}
  if even(n) -> goto D
   | odd(n) -> n := n-1; z := z+a; goto D
  fi
D: A & n>1 & even(n)
  n := n/2; a := a+a; goto B
H: z = n0*a0
\end{verbatim} 

\paragraph{Thoughts on the example}
An attractive feature of the {\tt if...fi} construct is
that it can contain any number of guarded commands.
This gives separation of concerns in several directions:
it allows correctness to be strictly partial for the time
being; it allows non-determinism for the time being.

Yet in this example there are always two guarded commands
without however being equivalent to
{\tt if...then...else...}
The rigid adherence to the binary {\tt if...fi}'s is a
consequence of the goal to obtain code that does not duplicate
tests.

\subsection{Example: Table of primes}
The most prominent example in Dijkstra's chapter
in \cite{daDiHo}
is to write a program that produces a table of the first
thousand prime numbers.
Dijkstra chose the method of trial division. Some
readers reacted with ``But everyone knows that the most
efficient way to generate prime numbers is by using the
Sieve of Eratosthenes'' (\cite{daDiHo}, page 27).

Dijkstra proceeds with an excruciatingly detailed account
of successive levels of refinement, until at the end
(page 37) he writes ``To give the algorithm an unexpected
turn we shall assume the absence of a convenient remainder
computation.''
He shows a simple way of avoiding the remainder operation,
which turns out to be the Sieve of Eratosthenes.

Abstinence from a built-in remainder operation comes at
a cost, namely, the need to create an auxiliary array
containing suitable multiples of as many of the smallest
primes as this array will hold.

I work by successive refinement by developing two
Trial Division and Sieve A.

Problem statement:
\begin{verbatim}
int p[1000]; int n := 1000;
S: int p[0..n-1] is allocated & n >= 2}
H: p[0..n-1] contains the first n prime numbers in incr. order }
\end{verbatim}
I can't think of any
verification condition with S as precondition
and H as postcondition.
Heuristic: if direct attack is not successful, then find
related easier problem that can be solved.
In this case, if not all of the $n$ prime numbers can be produced,
maybe $k$ ($k<n$) of them can.
Which $k$ primes? The $k$ largest? The $k$ smallest?
Last option is more plausible as it is likely that
knowing smaller ones can help in finding larger ones. 
Specifically, we can find the next larger prime
after the largest found so far.
\begin{verbatim}
int p[1000]; int n := 1000;
S: int p[0..n-1] is allocated & n >= 2
   p[0],p[1] := 2,3; k := 2; goto A
A: S & p[0..k-1] contains the first k prime numbers in incr. order
  if k = n -> goto H
   | k != n { 2 <= k < n } -> 
     "create variable cand, compute candidate for next prime 
      after p[k-1], and assign it to cand"
  fi
H: p[0..n-1] contains the first n prime numbers in incr. order
\end{verbatim}
How to find the next prime? Find an easier problem, one that is
easy to achieve initially and can be gradually nudged closer
to {\tt cand} being the next prime.
\begin{verbatim}
int p[1000]; int n := 1000;
S: int p[0..n-1] is allocated & n >= 2
   p[0],p[1] := 2,3; k := 2; goto A
A:  S & p[0..k-1] contains the first k prime numbers in incr. order
  if k = n -> goto H
   | k != n { 2 <= k < n } -> cand := p[k-1]+2; j := 0; goto B
  fi
B: A & cand not divisible by any of p[0..j]
  if cand < p[j]^2 { cand is next prime after p[k-1] } ->
     p[k] := cand; k := k+1; goto A
   | cand >= p[j]^2 -> 
     "set of possible values for cand needs to be shrunk"
  fi
H: p[0..n-1] contains the first n prime numbers in incr. order
\end{verbatim}
This leaves us with an untreated case of uncertainty 
about the primality of {\tt cand}.
\begin{verbatim}
int p[1000]; int n := 1000;
  int k, cand, j;
S: int p[0..n-1] is allocated & n >= 2
   p[0],p[1] := 2,3; k := 2; goto A
A: S & p[0..k-1] contains the first k prime numbers in incr. order
  if k = n -> goto H
   | k != n { 2 <= k < n } -> cand := p[k-1]+2; j := 0; goto B
  fi
B: A & cand not divisible by any of p[0..j]
  if cand < p[j]^2 { cand is next prime after p[k-1] } ->
     p[k] := cand; k := k+1; goto A
   | cand >= p[j]^2 -> j := j+1; goto C
  fi
C: A & cand not divisible by p[0..j-1]
  if cand%p[j] = 0 {cand not a prime} ->
     cand := cand+2; j := 0; goto B
   | not cand%p[j] = 0 -> goto B
  fi
H: p[0..n-1] contains the first n prime numbers in incr. order
\end{verbatim}
No sub-problem left unsolved. Ready for transcription to C.
The transcription leaves the structure of the Liffig code
intact, so that the verification conditions can be verified
in the C code. This completes Trial Division.

To obtain Sieve we need modifications at the
end of Trial Division.
A modification occurs at the beginning as well:
the declaration of the auxiliary array for the multiples of
the first few primes.
In Sieve A {\tt cand} is tested for divisibility by {\tt p[j]}
by comparing it with {\tt mult[j]}.
If equal, then cand is not prime.
Otherwise, and if {\tt mult[j] < cand}, we increase {\tt mult[j]}
by {\tt p[j]} and compare again.
This leads to the following program.
\begin{verbatim}
int p[1000]; int n := 1000;
int mult[30]; // mult[i] for multiple of p[i]
  int k, cand, j;
S: int p[0..n-1] is allocated & n >= 2
  p[0],p[1] := 2,3; k := 2; goto A
A: S & p[0..k-1] contains the first k prime numbers in incr. order
   k <= n
  if k = n -> goto H
   | k != n { 2 <= k < n } -> cand := p[k-1]+2; j := 0; goto B
  fi
B: A & cand not div by p[0..j]
  if cand < p[j]^2 { cand is next prime after p[k-1] } ->
     p[k] := cand; k := k+1; goto A
   | cand >= p[j]^2 -> j := j+1; mult[j] := p[j]; goto C
  fi
C: A & cand not div by p[0..j-1] & cand >= mult[j]
   if cand = mult[j] -> {cand is not a prime}
      cand += 2; j := 0; goto B
    | cand >  mult[j] -> mult[j] += p[j]; goto D
   fi
D: B & cand not div by p[0..j-1]
   if cand < mult[j] -> {cand not div by p[j]} goto B
    | cand >= mult[j] -> goto C
   fi
H: p[0..n-1] contains the first primes in increasing order
\end{verbatim}

\paragraph{Thoughts on the example}
In Trial Division we used refinement by first posting
informally the goal
\begin{verbatim}
"create variable cand, compute candidate
 for next prime after p[k-1], and assign it to cand"
\end{verbatim}
and then resolving it.
In the course of this resolution another refinement cropped
up in the form of
\begin{verbatim}
"set of possible values for cand needs to be shrunk"
\end{verbatim}

Distinct from the refinement technique is the strategy
of developing the least ambitious version, Trial Division,
first, and then finding a suitable modification to reach
Sieve.

Turning Trial Division into a possibly novel version of
the Sieve of Eratosthenes was a stroke of genius on the 
part of Dijkstra, which he passed by without remark.
He noted the difficulty that a size for the array of
multiples needed to be declared in advance. He wrote
`` number theory gives us 30 as a safe upper bound''.
Presumably that means invoking the Prime Number Theorem to
get an upper bound for the 1000-th prime, taking the
square root of it and invoke the Prime Number Theorem
again to conclude that the 30-th prime is a safe upper
bound for that square root.
I say, better make a guess at the size of the {\tt mult}
array and build an emergency exit into the code  for the
situation that this size turns out to be too small.
Or switch to C++ and use the standard library's {\tt vector}
container to store the multiples.
Neither of these low-level maneuvers detract from Dijkstra's
brilliant idea.

\section{{\large Concluding remarks}}
\paragraph{From Matrix Code to Liffig}
Matrix code was born on a whiteboard.
As a result it was little used.
To be usable it is necessary to transform it to a format
that is easy to store in a computer file and easy to edit
with a keyboard.
The key to such a format is the observation that code matrices
are sparse, which suggests as alternative to the two-dimensional
matrix format a list of triples consisting of column label,
cell content, and row label.
This is the list of verification conditions.
The matrix format suggests ordering the triples
by column label and using the {\tt if...fi} of Dijkstra's
guarded command language.
The desirability of verification by assertions suggests the use
of labels to mark locations in the code where the assertions
are intended to hold.
It also suggests a formal way of claiming that a certain
assertion holds that has already been labelled.
For this the keyword {\tt goto} is an obvious choice,
as {\sl of course} it has no other use.
These observations amount to a description of the programming language
Liffig.

\paragraph{What counts as a proof?}
In the 1930's predicate logic stabilized into the currently
conventional syntax and semantics.
From the same decade stems a new standard of rigour in mathematics
as exemplified by van der Waerden's ``Moderne Algebra''
and the first fascicles produced by Bourbaki.
Mathematicians have accepted as relaxed notion
of rigour that in principle
a formalisation in first-order predicate logic
is possible, but that in practice an informal summary
is preferable.
I have adopted this relaxed notion of rigour.

When, for example in the verification conditions for the
program for the table of primes I write
\begin{verbatim}
cand not div by p[0..j-1]
\end{verbatim}
I mean the formula
$$
\forall x.\,(0 \leq x \wedge x \leq j-1
\Rightarrow \neg div(cand, p[x])).
$$
which is still problematic from the point of view of
predicate logic because of the ``variable'' $p[x]$.

\paragraph{Need for an expanded version of predicate logic for
assertions}
In Section~\ref{sec:floydLogic}
Floyd's logic was defined to include predicate logic as the language
for the assertions.
This is not adequate for most programs where variables have
sorts (``types'')
and are organized in arrays and other data structures.
Proposals for such an extension are presented
in Apt and Bezem \cite{aptBezem}.

\paragraph{What is imperative programming?}
Until recently I recognized two programming paradigms:
imperative versus declarative.
Imperative is characterized by being dominated by the concept of
state, whereas state plays no role in declarative programming.
Indeed, although Figure~\ref{prog:spaghetti} is based on the
equations in Figure~\ref{tex:stein}, the relation is not close.

Lamport \cite{lamport08} has convinced me that the dichotomy
of imperative versus declarative
sketched above is not helpful.
A program in Scheme or in Prolog, though declarative,
still specifies a computation
and according to Lamport all computation is performed by
a state machine of some sort. This is obviously true of
imperative program, but it is also true of programs in Scheme
or Prolog, although typically less obviously so.

Still, structured programming is different from the kind of programming
for which Scheme or Prolog is intended for. How can one characterize
the difference? Not by whether a state machine is involved. The
difference seems to lie in the structure of the state and in its
visibility. In the kind of programming that Dijkstra is interested
in, and that I loosely refer to as imperative programming, the
structure of the state is simple: the state is a vector of primitive
values indexed by variables. The state is visible: new values are
created as values of expressions containing state components and
decisions are taken on the basis of such expressions. In the kind
of programming for which Scheme and Prolog are intended, the structure
of the state is complex, containing stacks and pointers into them.
These are not visible and are manipulated indirectly via the
evaluation of expressions (Scheme) or via the elaboration of goal
statements (Prolog).

\paragraph{``Goto Considered Harmful''}
Although the term ``structured programming'' does not occur in
Dijkstra's 1968 Letter to the Editor \cite{dijkstra68}, the two
have been strongly associated by the public.
In the Letter Dijkstra advocates
abolishing the goto statement in any programming language above
the level of machine language and to use structured control
exclusively. I end up using the goto statement exclusively.
There must be something in the Letter that I disagree with.
What is it?

Dijkstra opens the argument with
\begin{quote}
My first remark is that, although the programmer's activity ends
when he has constructed a correct program, the process taking place
under control of his program is the true subject matter of his
activity, for it is this process that has to accomplish the desired
effect; it is this process that in its dynamic behaviour has to
satisfy the desired specifications.
\end{quote}
What I propose is to change
``the true subject matter of his activity'' from that process to the
{\sl goal} the process is to achieve. That goal is a static entity. It
can be expressed as an assertion. The programmer's activity should
be to design achievable goals and to link their assertions with
verification conditions. One can statically ascertain whether a set
of verification conditions is sufficient for the task at hand.

To
proceed thus is to work at a higher level than structured programming;
if it needs a name, then ``goal-directed programming'' might do. From
the point of view of \cite{dijkstra71}
this is a step forward because it is assumed
there that code has to be verified. Indeed goal-directed programming
results in code that is verified in the sense of Floyd. But I see
a more important advantage. Without knowing an algorithm to achieve
the goal, the subgoals needed suggest themselves, together with
verification conditions that link them to a complete set. As observed
before, to get from a set of verification conditions to executable
code is trivial mechanics. 
The examples in this paper demonstrate this miracle.

\appendix

\section*{Acknowledgements}
In writing this paper I received much support in various forms
from Paul McJones.

\bibliographystyle{plain}
\bibliography{bibfile}

\end{document}